\documentclass[aip,jap,reprint,groupedaddress,superscriptaddress,floatfix]{revtex4-1}
\usepackage{ulem}
\usepackage{amsmath}
\usepackage{xcolor}

\newcommand{\diff}{\mathrm{d}}

\newcommand{\Ef}{{ E_{\rm F}}}

\usepackage{units}
\usepackage{bm}
\usepackage{hyperref}
\usepackage[T1]{fontenc}
\usepackage[latin1]{inputenc}
\usepackage{amssymb}
\usepackage{amsmath}
\usepackage{amsfonts}
\usepackage{color}
\usepackage{graphicx}
\usepackage{natbib}

\begin{document}
\title{Shallow impurity band in ZrNiSn} 

\author{Matthias Schrade}
\email{matthias.schrade@sintef.no}
\affiliation{Department of Physics, Centre for Materials Science and Nanotechnology, University of Oslo, Sem S{\ae}landsvei 26, 0371 Oslo, Norway}
\affiliation{SINTEF Materials Physics, Forskningsveien 1, 0373 Oslo, Norway}
\author{Kristian Berland}
\email{kristian.berland@nmbu.no}
\affiliation{Department of Physics, Centre for Materials Science and Nanotechnology, University of Oslo, Sem S{\ae}landsvei 26, 0371 Oslo, Norway}
\affiliation{Faculty of Science and Technology, Norwegian University of Life Sciences, 1433 \AA s, Norway}
\author{Andrey Kosinskiy}
\affiliation{Department of Physics, Centre for Materials Science and Nanotechnology, University of Oslo, Sem S{\ae}landsvei 26, 0371 Oslo, Norway}
\author{Joseph P. Heremans}
\affiliation{Department of Mechanical and Aerospace Engineering, Ohio State University, Columbus, Ohio 43210, USA}
\affiliation{Department of Physics, Ohio State University, Columbus, Ohio 43210, USA}
\author{Terje G. Finstad}
\affiliation{Department of Physics, Centre for Materials Science and Nanotechnology, University of Oslo, Sem S{\ae}landsvei 26, 0371 Oslo, Norway}

\date{\today}

\begin{abstract}
ZrNiSn and related half Heusler compounds are candidate materials for efficient thermoelectric energy conversion with a reported thermoelectric figure-of-merit of n-type ZrNiSn exceeding unity. Progress on p-type materials has been more limited, which has been attributed to the presence of an impurity band, possibly related to the presence of Ni interstitials in nominally vacant $4d$ position.
The specific energetic position of this band, however, has not been resolved. 
Here, we report results of a concerted theory-experiment investigation for a nominally undoped ZrNiSn, based on measurements of electrical resistivity, Hall coefficient, Seebeck coefficient and Nernst coefficient, measured in a temperature range from 80 to \unit[420]{K}. The results are analyzed with a semi-analytical model combining a density functional theory (DFT) description for ideal ZrNiSn, with a simple analytical correction for the impurity band.
The model provides a good quantitative agreement with experiment, describing all salient features in the full temperature span for the Hall, conductivity, and Seebeck measurements, while also reproducing key trends in the Nernst results. 
This comparison pinpoints the impurity band edge to \unit[40]{meV} below the conduction band edge, which agrees well with a separate DFT study of a supercell containing Ni interstitials. Moreover, we corroborate our result with a separate study of ZrNiSn$_{0.9}$Pb$_{0.1}$ sample showing similar agreement with an impurity band edge shifted to \unit[32]{meV} below the conduction band.
\end{abstract}

\maketitle

\section{Introduction}
\label{Introduction}
Materials which combine environmental abundance and low toxicity with good thermoelectric properties 
are as rare as they are sought after. 
Notable exceptions are the half Heusler compounds $X$NiSn and $X$CoSb with ($X=$ Hf, Zr, Ti), 
which combine good thermoelectric performance for n-doped samples
with chemical stability in the mid to high temperature range from 400 to \unit[900]{K}.\cite{Xie2012b} 

For $n$-type $X$NiSn, the thermoelectric figure-of-merit $zT$,\cite{Goldsmid2017} has been reported to exceed unity in a wide temperature range,\cite{Sakurada2005,Schwall2013,Joshi2011,Guerth2016} while experimental efforts to p-dope $X$NiSn-based materials have only lead to modest $zT$ values below 0.1.\cite{Schmitt2015,Xie2012c,Kimura2010}
The difficulties in p-doping these materials have been linked to 
the presence of Ni interstitials
in $X$NiSn giving rise to impurity band states within the band gap.\cite{Zeier2016,Tang2018,Downie2015a}

The concentration of Ni interstitials varies with fabrication method and sintering temperature, with reported values between 1 and \unit[8]{\%}.\cite{Tang2018,Barczak2018,Xie2012,Guzik2018,Miyazaki2014,Hu2018,Downie2015}
Angle-resolved photoemission spectroscopy on single-crystals\cite{Fu2019} and DFT-based calculations\cite{Silebarski1998,Ouardi2011} of pure ZrNiSn agree on an intrinsic band gap of around \unit[0.5]{eV}, whereas optical measurements on polycrystalline samples show an absorption onset at much lower values, around $E_g=\unit[0.13]{eV}$, possibly related to impurity states within the band gap.\cite{Schmitt2015}
However, if this optical band gap coincided with the upper edge of the impurity band, it would require an immense concentration of interstitial Ni,
in order to explain typical values for the intrinsic charge carrier concentration found for these compounds.
For example, Xie {\itshape et al.}\cite{Xie2014} 
reported a carrier concentration of $n_{\rm intrinsic} = \unit[5\cdot10^{19}]{cm^{-3}}$ at \unit[300]{K} for nominally undoped ZrNiSn.
Assuming this conduction band carrier concentration was primarily due to thermal excitation from impurity states located \unit[0.13]{eV} below the conduction band edge, 
we obtain a donor concentration 
of $N_D\approx \unit[8\cdot10^{21}]{cm^{-3}}$. 
Assuming at most four available electrons per Ni atom, corresponding to the $e_g$ states of interstitial Ni,\cite{Do2014} this value would correspond to an unphysically high Ni interstitial concentration of at least \unit[50]{\%}.

Instead of Ni interstitials, there are other possible scenarios for the formation of in-gap states and - if the density is high enough - their hybridization into an impurity band: Structural defects like dislocations or other types of atomic disorder, as for example anti-site defects between the Zr and Sn sublattice, all break the periodicity of the crystal and can modify the band structure. Indeed, an early study on ZrNiSn reported a high concentration of Zr/Sn anti-site defects of \unit[30]{\%},\cite{Aliev1987} while later, more detailed work found anti-site defects rather unlikely in this material, with Ni interstitials being the dominating defect.\cite{Xie2012} Also a computational study found the lowest formation energy and thus the highest defect concentration for Ni interstitials in ZrNiSn,\cite{Miyazaki2014} and we will thus discuss our results under this assumption. However, we emphasize that it is not the goal of the current paper to pinpoint the microscopic origin of the impurity band, and that the obtained characteristics of the impurity band are independent of its specific origin.

In this paper, we investigated the electronic transport properties of ZrNiSn and the closely related ZrNiSn$_{0.9}$Pb$_{0.1}$ by measuring the resistivity, the Seebeck coefficient, Hall coefficient and Nernst coefficient. Substituting Sn by Pb was recently proposed as an efficient way to reduce the thermal conductivity, avoiding expensive Hf.\cite{Mao2017} Our experimental data was analyzed with a model combining input from DFT calculations with an analytical correction describing
the presence of an impurity band. The model exhibits excellent agreement with experimental data. Our model provides estimates of key properties such as the
energetic position of the top of the impurity band, the mobility of associated states, and the order of magnitude of the number of states in the impurity band.

\section{Methods}
\subsection{Experimental details}
Two samples with nominal composition, ZrNiSn and ZrNiSn$_{0.9}$Pb$_{0.1}$, were prepared via an arc-melting, crushing, annealing, and sintering routine, as described in Ref.~\onlinecite{Kosinskiy2016}: Stoichiometric amounts of metallic pieces of Zr, Ni, and Sn (purity 99.9 wt.\% or higher) were arc-melted in a Ti-gettered Argon-atmosphere. Samples were turned and remelted several times to increase homogeneity. Resulting samples were crushed and ball-milled in an argon atmosphere.  For the ZrNiSn$_{0.9}$Pb$_{0.1}$ sample, pieces of Pb were then added to the powder. The two mixtures were then annealed in vacuum-closed silica ampoules for 40 days at \unit[1123]{K}. The resulting samples were ballmilled a second time and then sintered at \unit[1273]{K} for \unit[10]{min} under an uniaxial pressure of \unit[60]{MPa} using an in-house made hot-press. Phase composition and microstructural properties were screened during synthesis using X-ray diffraction and scanning electron microscopy.  For ZrNiSn, we obtained a homogeneous, single phase pellet with an almost stoichiometric composition, as characterized with energy dispersive spectroscopy (EDS). The ZrNiSn$_{0.9}$Pb$_{0.1}$ sample phase separated into two main phases, both with a half Heusler symmetry and a chemical composition of ZrNiSn$_{0.88}$Pb$_{0.12}$ and ZrNiSn$_{0.94}$Pb$_{0.06}$ by EDS, Figs. S1--S4. Additional annealing did not significantly modify sample homogeniety.
The electrical resistivity and the Hall and Seebeck coefficient of the polycrystalline samples were measured from 80 to \unit[420]{K} using a steady-state four point heater-and-sink method.\cite{Heremans2004}

\subsection{Computational details}
The DFT calculations made use of the \textsc{VASP} \cite{vasp1,vasp3,vasp4} software package. All structure relaxations were performed with the PBEsol functional\cite{PBEsol},
as it generally provides more accurate lattice constants\cite{csonka_assessing_2009} than   PBE.\cite{pbe1996}
Transport properties of ZrNiSn were calculated similar to that of Ref.~\onlinecite{berland:group4},
combining the \textsc{BoltzTraP}\cite{boltztrap} software package and a recently developed $\bm{k} \cdot \bm{p}$-based interpolation method,\cite{berland2018_kp_thermo} interpolating to a $60\times60\times60 $ $\bm{k}$-mesh. The input electronic structure was computed at the generalized-gradient level as in  Ref.~\onlinecite{berland:group4} rather that at the hybrid functional level, as we found the former to generally agree better with the measured room temperature Seebeck coefficients of Xie~{\itshape et al.}\cite{Xie2014} at different doping concentrations.

The electronic properties of ZrNiSn$_{0.9}$Pb$_{0.1}$ was modeled with an expanded lattice parameter of \unit[6.0786]{\AA}, linearly interpolating the relaxed unit cells of ZrNiSn ($a=\unit[6.0702]{\text{\AA}}$) and ZrNiPb ($a=\unit[6.1514]{\text{\AA}}$), adapting the same procedure as Bhattacharya {\itshape{et al.}}.\cite{Bhattacharya2015} This approach assumes the validity of Vegard's law for this material system, as has been experimentally confirmed by Mao {\itshape et al.} \cite{Mao2017}. The DFT-computed lattice parameters are very close to reported experimental values,\cite{Gautier2015,Mao2017,Xie2012} with deviations of \unit[-0.7]{\%} and \unit[+0.5]{\%} for, respectively, ZrNiSn and ZrNiPb.
The validity of the volumetric-expansion approach was also tested by replacing Pb by Sn for the ZrNiPb crystal structure, which resulted in virtually identical transport properties for a given relaxation time $\tau$.
Supplementary supercell DFT calculations were also performed with  $2\times2\times2$ cubic unit cell, with one nickel atom in the supercell corresponding a Ni occupation of approximately 3\%. 

\section{Experimental results}
\begin{figure}[htbp]
\centering
\includegraphics[width=0.38\textwidth]{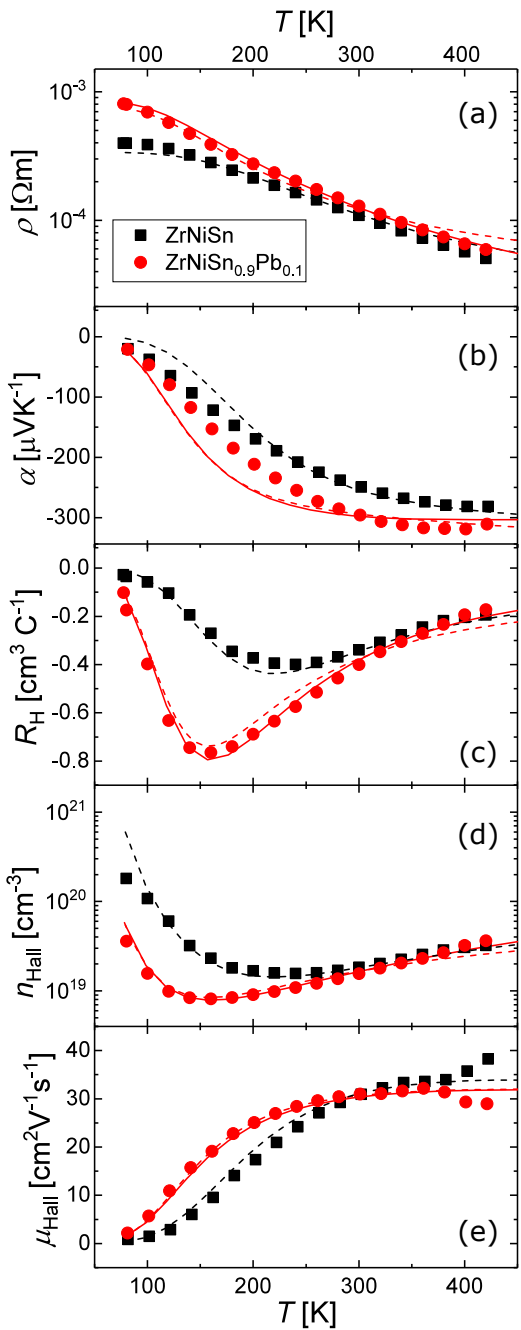}
\caption{Electrical resistivity (a), Seebeck coefficient (b), Hall coefficient (c), Hall concentration, $n_H=1/R_He$ (d), and Hall mobility (e) of ZrNiSn (black) and ZrNiSn$_{0.9}$Pb$_{0.1}$ (red). The dashed curves indicate the results from the presented model, while in the full curves an extra deep carrier reservoir is included.}
\label{fig_raw_data}
\end{figure}
Fig.~\ref{fig_raw_data} displays obtained experimental data in qualitative agreement with data by Uher {\itshape et al.}\cite{Uher1999}  
Substituting a fraction of Sn with Pb does not change the overall shape of the measured curves, as is expected for an isoelectronic substitution. The observed variation between ZrNiSn and ZrNiSn$_{0.9}$Pb$_{0.1}$ can rather be related to different levels of unintended impurities, as discussed in section \ref{discussion}. 

The experimental data shows the following salient features:
\begin{description}
  \item[Vanishing $\alpha$ at low temperature (\ref{fig_raw_data}b)]
This is consistent with Fermi level pinning within the impurity band, i.e. a characteristic of a partly occupied impurity band.
\item[A broad miminum in $\alpha$ (\ref{fig_raw_data}b)]
The absolute value of the Seebeck coefficient increases with increasing temperature until it reaches a broad maximum around \unit[400]{K}. Such results are usually associated with the onset of bipolar conduction, but could also arise from increased carrier concentrations in the conduction band.  
\item[A minimum in $R_H$ (\ref{fig_raw_data}c)]
An unusual decrease of $\left|R_H\right|$ with decreasing temperature has also been observed earlier for $X$NiSn compounds.\cite{Galazka2014,Berry2017,Hu2018,Romaka2013}
In general, it is difficult to interpret $R_H$ in terms of the carrier concentration due to multiband behavior or complicated shapes of the Fermi surface.\cite{Ong1991}
An extremum in the Hall coefficient $R_H$ is usually associated with two bands contributing in a similar amount to the transport.\cite{Hung1954,Pei2016} 
This can be understood in terms of a two band model, with $R_H$ given by
\begin{align}
    R_H = \frac{n_1\mu_1^2 \pm n_2 \mu_2^2 }{e(n_1\mu_1 + n_2 \mu_2 )^2}
  \label{eq:Rh}
\end{align}
where $n_{1(2)}$ and $\mu_{1(2)}$ are, respectively, the concentration and mobility of charge carriers in band 1(2). 
The plus sign is used if the conduction in both bands is of the same type (electrons or holes) and the mignus sign for the opposite case.
Assuming $\mu_1 >> \mu_2$ -- which would generally be the case for an impurity bands -- $R_H$ has a minimum at $n_1 \mu_1 = n_2 \mu_2$.
\end{description}

\section{Theoretical model and discussion}
\label{discussion}
In the following, we describe the model used to analyze our experimental results.
Our model combines a DFT description of the valence and conduction band of ZrNiSn with a rougher analytical description of the impurity band.
The reason for using DFT for the these bands is that it automatically includes band degeneracies and non-parabolicity without the need for additional empirical parameters, such as an effective conduction band mass. 

The transport properties are calculated in the Boltzmann transport equation (BTE).\cite{boltztrap}
For a given Fermi level $\Ef$ the thermoelectric transport contributions can be calculated from the obtained density of states (DOS) $g_0(\epsilon)$, the transport DOS, $\Sigma(\epsilon)$\footnote{Note that our definition of $\Sigma(\epsilon)$ deviates from the expression as defined by Madsen {\itshape et al.}\cite{boltztrap} by the constant factor $e^2\tau$}, 
and Hall transport DOS $\Sigma_{\text{,H}}(\epsilon)$ in terms of the derivative of Fermi-Dirac function,
$  f_1(\beta \epsilon) =  -1/\beta \partial f_{\rm FD}(\epsilon)/\partial \epsilon  = \left[\exp(\beta \epsilon) + 2 + \exp(-\beta \epsilon)\right]^{-1}$,
as follows
\begin{align}
\sigma &= e^2 \beta  \int_{-\infty}^\infty \diff \epsilon\, \tau(\epsilon) f_1[\beta (\epsilon -\Ef)] \Sigma(\epsilon) \,,  \label{eq:sigma_Sig} \\
T\sigma \alpha &= e   \beta  \int_{-\infty}^\infty \diff \epsilon\,   \tau(\epsilon) f_1[\beta (\epsilon -\Ef)]  \Sigma(\epsilon) (\epsilon-\Ef) \,, \label{eq:alpha_Sig}\\
\sigma^2 R_{\rm H} &=     \beta \int_{-\infty}^\infty \diff \epsilon\, \tau(\epsilon) f_1[\beta (\epsilon -\Ef)]  \Sigma_{{\rm H}}(\epsilon) (\epsilon-\Ef)^2\,,\label{eq:Rh_Sig}
\end{align}
where $\sigma$, $\alpha$, $R_{\rm H}$ are the electrical conductivity, Seebeck coefficient, and Hall coefficient.
The contributions to the density of states $g_0(\epsilon)$ and transport spectral function $\Sigma_0(E)$  for valence and conduction band states were computed using the \textsc{BoltzTraP} package.\cite{boltztrap} 
To limit the number of adjustable parameters,
we employed a single constant relaxation time $\tau$ for valence and conduction band states. While using the same relaxation time for the conduction and valence
is not realistic, this is inconsequential, as we find no appreciable contributions from the valence band to the transport properties at the studied temperatures. 

In our model, we neglect broadening of the impurity band, so that the full DOS is approximated

by $g(E) = g_0(E) + N_p \delta(E - E_p)$, where $N_{\rm p}$ is the impurity band density and $E_{\rm p}$ is the energetic position of the impurity band,
and correspondingly for the full transport spectral function,
$\Sigma(\epsilon)=\Sigma_0(\epsilon)+e N_{\rm p} \mu_{\rm p} \delta(\epsilon -  E_{\rm p})\,$. This expression can account for the possibility of transport in the impurity band channel, where $\mu_p$ is the impurity band mobility. 
In terms of the description of valence and conduction bands\cite{boltztrap}, this mobility can be understood as the mean $\langle v_g^2 \tau_p \rangle \propto  \langle \mu_p \rangle$, where $v_g$ is the impurity band group velocity and $\tau_p$ is the corresponding relaxation time, evaluated prior to taking the limit of vanishing bandwidth. 
We also ignore any indirect effects an impurity band could have on the conduction band dispersion, beyond introducing scattering of the conduction band electrons, which is accounted for by the adjustable relaxation time $\tau$. 

In the full model, the conductivity is given by 
\begin{align} 
    \sigma &= \sigma_0 +   e N_{\rm p} \mu_{\rm p} f_1[\beta( E_{\rm p} - \Ef )],
\label{eq:cond_imp}
\end{align} 
and the corresponding generalization of Eq.\,\eqref{eq:alpha_Sig} is given by
\begin{align}
T \sigma \alpha   &=  T \sigma_0 \alpha_0  +   \beta N_{\rm p} \mu_{\rm p}  f_1[\beta (E_{\rm p} - \Ef)]  (E_{\rm p} -\Ef). \label{eq:sig_imp}
\end{align}
Finally, the full Hall coefficient is given by 
\begin{align}
\sigma^2 R_H  =  R_H\sigma_0^2+ e N_{\rm p} \left[\mu_{\rm p}^2 f_1(\beta( E_{\rm p} - \Ef )) \right]. \label{eq:hall_imp}
\end{align}

Key approximations in this model, is the use of  i) a fixed temperature-independent relaxation time for the conduction and valcence band, ii) a fixed temperature-independent mobility for the impurity band, iii) vanishing bandwidth for the impurity band. 
Clearly, these are coarse approximations, but that are chosen to keep the number of adjustable parameters to a minimum. 
The lack of bandwidth should also be viewed as only describing the upper part of the impurity band, 
which is the only part that is critical to include in the model:
Once a sufficient number of the electrons originating in the impurity band
has been excited to the conduction band, the transport is in any case dominated by conduction band transport,
and the exact value of the impurity band mobility is less critical for the the overall transport properties.
By the same reasoning, and as a numerical convenience to limit the number of adjustable parameters, we will assume that
the conductive impurity band channel
is half filled at $T=\unit[0]{K}$. The lack of full occupancy is attributed to  acceptor levels located deep in the band gap and not contributing to the transport properties.
The essential mechanism of half-filling is to prevent a fully occupied impurity band even at the lowest temperatures.

\begin{table}[tb]
\caption{The optimized parameters of the impurity band model used to analyze the experimental data.}
 \begin{tabular}{lccc}
\hline
Parameter & ZrNiSn & ZrNiSn$_{0.9}$Pb$_{0.1}$ \\
\hline
$\tau $ [10$^{-14}$s] &  1.85 & 1.37  \\
$\Delta = E_{\rm C} - E_p $ [meV] & 40 & 32\\
$N_{p}$ [10$^{19}$cm$^{-3}$] & 18  & 6 \\
$\mu_{p}$ [cm$^2$V$^{-1}$s$^{-1}$] & 4.1 & 4.9\\
\hline
$N_{\rm res}$ [10$^{19}$cm$^{-3}$] & - & 20\\
$E_{\rm res}$ [{\rm eV}] & - & 0.13\\
\hline
\end{tabular}
 \label{tab_fitting_parameters}
\end{table}

A shortcoming of the model, is that it can not describe additional band filling of the conduction band once the upper part of the conduction band is emptied. We therefore also explored models using a second electron reservoir with $N_{\rm res}$ states located at $E_{\rm res}$.
For ZrNiSn, we found no appreciable improvement of the fit with an extra reservoir band, so we set $N_{\rm res} = 0$,
as one p-channel impurity band was sufficient to quantitatively describe the measured thermoelectric transport properties.
In the case of the ZrNiSn$_{0.9}$Pb$_{0.1}$ sample, on the other hand,
which has less states in the p-conductive channel, 
for the temperature range beyond \unit[300]{K}, introducing an extra reservoir level located \unit[0.13]{eV} below the band edge,
improved agreement with experimental data.
Incidentally, this corresponds to the earlier reported optical gap of ZrNiSn.\cite{Schmitt2015} 
The best fit of the model to the experimental data is shown in Fig. \ref{fig_raw_data} as solid and dashed lines, while the optimized parameters are summarized in Table \ref{tab_fitting_parameters}.
The model shows excellent agreement with all measured transport properties.

\begin{figure}[htbp]
\centering
\includegraphics[width=0.38\textwidth]{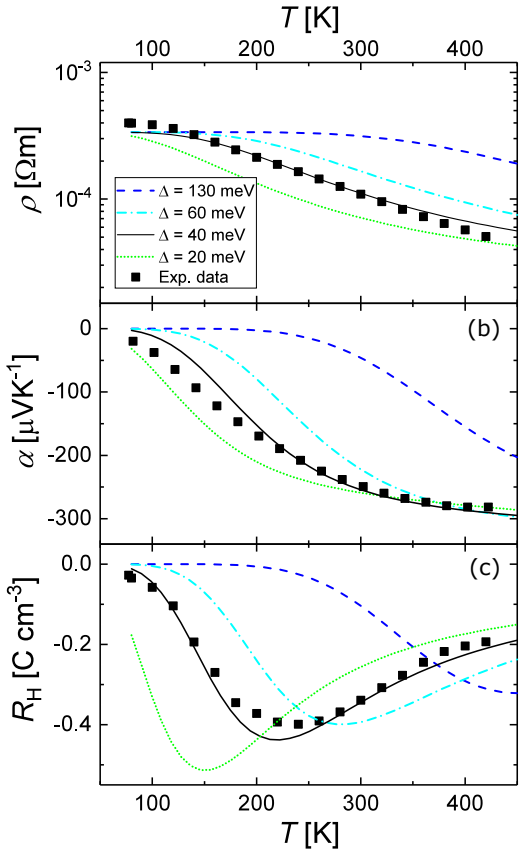}
\caption{Calculated resistivity (a), Seebeck coefficient (b) and Hall coefficient (c) for different positions of the impurity band. It is futile to obtain good agreement for all three curves for a value of $\Delta$ differing significantly from \unit[40]{meV}.}
    \label{fig_gap}
\end{figure}

We note that while there are slightly different combinations of $\mu_{\rm p}$, $\tau$, and $N_{\rm p}$ that all could provide 
relative good matches to the experimental data, 
the fit is very sensitive to the energetic position of the impurity band,
and only an impurity band \unit[$\sim 40$]{meV} below the conduction band edge can provide a good fit for 
the minimum in Hall carrier concentration, and the shape of the Seebeck and Hall mobility, as illustrated in Fig. \ref{fig_gap}.
Here, the other parameters are kept fixed to those optimized for $\Delta = \unit[40]{meV}$.

\begin{figure}[tbp]
\centering
	\includegraphics[width=0.38\textwidth]{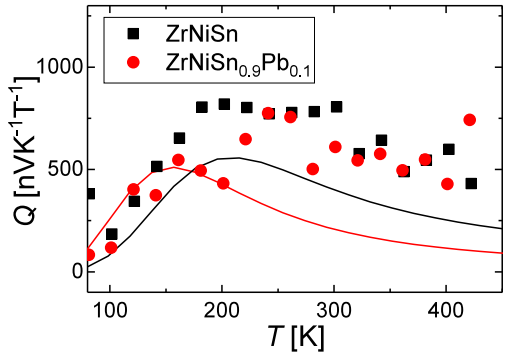}
	\caption{The Nernst coefficient $Q$ as a function of temperature. $Q$ shows small, positive values for both studied compositions. Solid curves show calculated results for the third term in Eq. \eqref{nernst_two_band} as described in the text.}
	\label{fig_nernst}
\end{figure}
\begin{figure}[htbp]
\centering
	\includegraphics[width=0.45\textwidth]{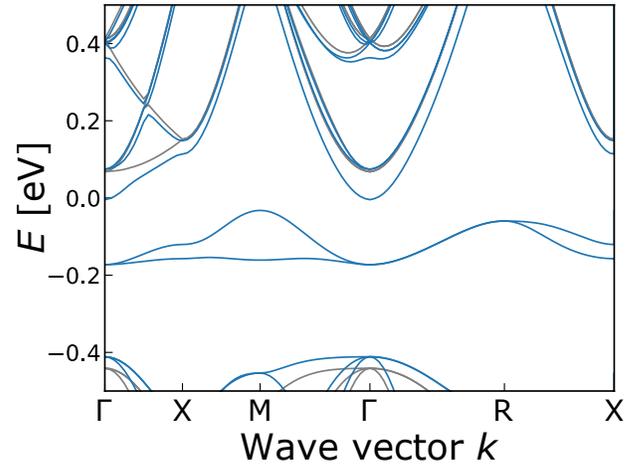}
	\caption{The band structure of a cubic $2\times 2\times 2$ supercell of ZrNiSn (grey) compared to the same supercell with an extra Ni interstial (blue). Upon Ni addition, an impurity band appears close to the conduction band edge.}
	\label{fig_supercell}
\end{figure}
\begin{figure}[htbp]
\centering
	\includegraphics[width=0.38\textwidth]{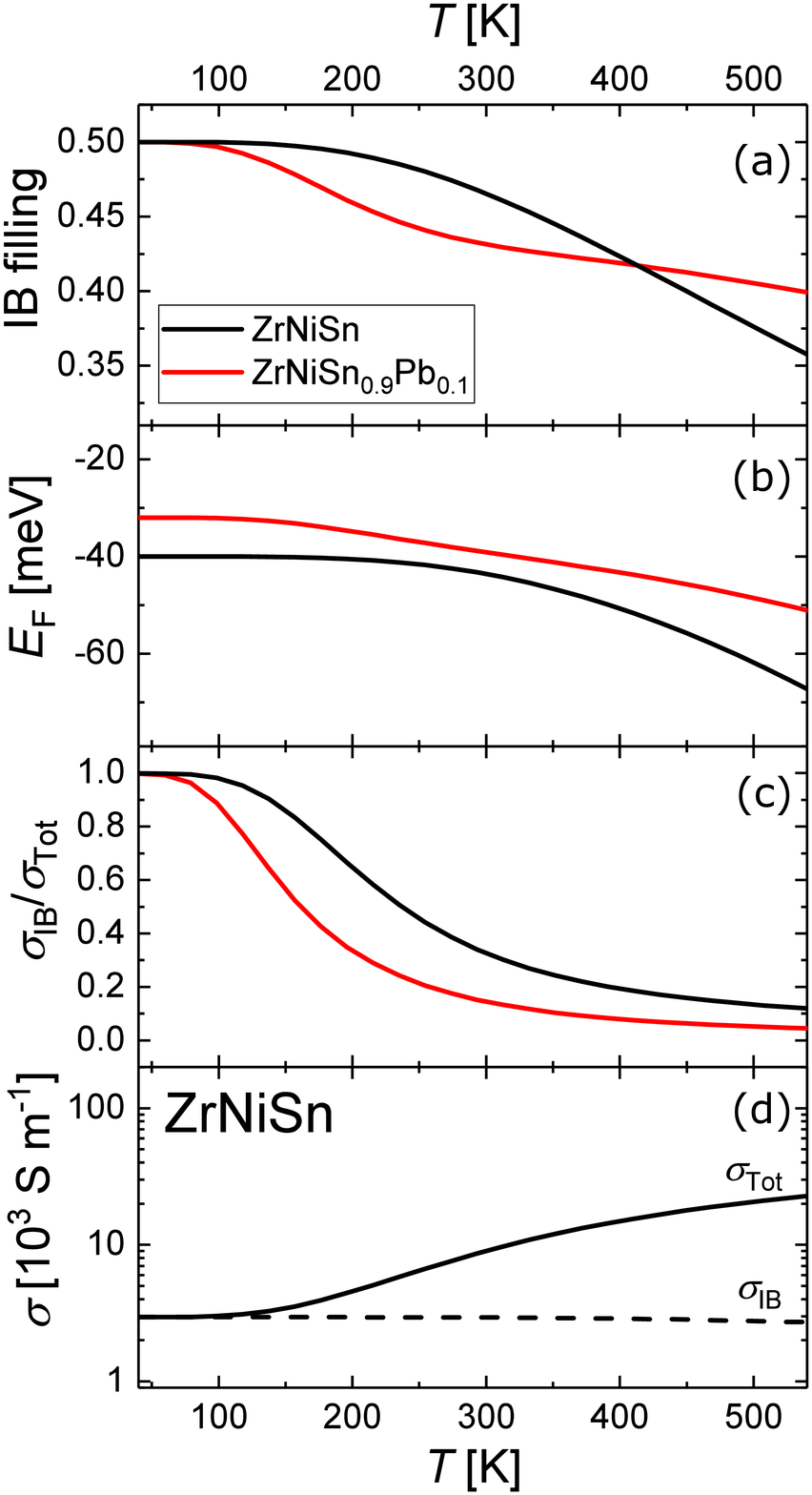}
	\caption{(a) Band filling of the impurity band as a function of temperature. (b) Calculated Fermi level for the two compositions. (c) Contribution of the impurity band to the total conductivity in the model. Low temperature transport is dominated by the impurity band, and even at room temperature $\sigma_{\text{IB}}$ accounts for \unit[$\sim30$]{\%} of the total conductivity. (d) Temperature dependence of $\sigma_\text{Tot}$ and $\sigma_{\text{IB}}$ of ZrNiSn.}
	\label{fig_impurity_conc}
\end{figure}

In addition to $\rho$, $\alpha$ and $R_H$, we also measured the Nernst coefficient $Q$ of ZrNiSn$_{1-x}$Pb$_x$. The Nernst effect is the  generation of a transverse electrical field by a longitudinal thermal gradient in the presence of a finite magnetic field $B$. The Nernst coefficient $Q$ can be expressed by the Hall angle $\theta_H$:\cite{Behnia2009}
\begin{equation}
Q=-\frac{\pi^2k_B^2T}{3eB}\left.\frac{\text{d}\tan{\theta_H}}{\text{d}\epsilon}\right|_{\epsilon=E_F}
\end{equation}
The Nernst coefficient is sensitive to scattering processes of the electrons contributing to charge transport: For example, for a simple, single band conductor, $\tan{\theta_H}$ is proportional to the scattering time $\tau$.\cite{Sun2013} For ZrNiSn$_{1-x}$Pb$_x$, $Q$ is positive, with much lower absolute values than the conventional Seebeck coefficient (Fig. \ref{fig_nernst}). Neither the temperature dependence, nor the absolute value of $Q$ seem to be affected by the Pb concentration.

For a two band system, the Nernst coefficient can be expressed as:\cite{Nemov1998}
\begin{equation}
Q=Q_\text{IB}+Q_0+(\alpha_\text{IB}-\alpha_0)(R_{H,\text{IB}}\sigma_\text{IB}-R_{H,0}\sigma_0)\sigma_\text{IB}\sigma_0/\sigma_\text{Tot}^2
\label{nernst_two_band}
\end{equation}
Here, $Q_i$, $\sigma_i$, $\alpha_i$ and $R_{H,i}$ are the Nernst coefficient, the conductivity, the Seebeck coefficient and Hall coefficient of electrons in the $i$th band. Often, the third, cross-term in Eq. \eqref{nernst_two_band} dominates over $Q_i$,\cite{Kajikawa2016}, and we therefore evaluate it for our model. The results are indicated as solid lines in Fig. \ref{fig_nernst}. Given the simplicity of our model, the overall agreement between experimental values and the cross-term is deemed good: At low temperatures, where the impurity band dominates the electronic properties, the calculated curves describe the experimental values excellently, while at higher temperatures, the calculated curves decay faster than the experimental values. This difference can be attributed to the intrinsic Nernst coefficient $Q_0$ of the conduction band.

As discussed in Section \ref{Introduction}, it is widely accepted that the impurity band in $X$NiSn compounds is related to the presence of Ni interstitials. Therefore, we have calculated the band structure of a $2\times2\times2$ supercell with one extra Ni occupying the nominally vacant $4d$ site, corresponding to an average Ni interstitial concentration of \unit[$\approx3$]{\%} with results shown in Fig. \ref{fig_supercell}.
Similar to the finding of Fiedler {\itshape et al.}\cite{fiedler2016ternary} and Do {\itshape et al.},\cite{Do2014} our DFT supercell calculations show the appearance of additional states just below the conduction band edge,
with $\Delta=\unit[18]{meV}$ for ZrNiSn  and \unit[16-28]{meV} for ZrNiSn$_{0.9}$Pb$_{0.1}$. Although these results agree well our finding of a shallow impurity band, caution is needed in interpreting these results and the band structure in Fig. \ref{fig_supercell}.
As a supercell is used, the bands are folded, making it appears as even pristine ZrNiSn (grey curves) has a direct band gap with triple degenerate conduction band minimum, while this is not the case for the FCC primitive cell, for which there is nondegenerate conduction band minimum at the $X$-point and a valence band maximum at the $\Gamma$-point. 
Second, modelling disordered, defective structures using ordered supercells introduces computational artefacts. Indeed, we observe a splitting of the conduction band for our supercell calculation as compared to pristine ZrNiSn. Such splitting cannot occur for fully disordered system containing Ni-interstitial as the conduction band minimum at the $X$-valley of FCC supercell is not degenerate.
However, the splitting does imply that Ni interstial strongly scatter the conduction band electrons making the exact position of the conduction band more diffuse, which could also contribute to explaining the presence of a shallow impurity band edge. 
The impurity band also exhibit a significant bandwidth. This bandwidth would be significantly smaller for a disordered system. 
Overall, while the DFT calculations are consistent with the existence of shallow impurity band, which could arise in part due to a renormalization of the conducting band itself, we hope our study will trigger in-depth DFT studies that can elucidate the exact role of the Ni interstitials. Such studies would demand 
bigger supercells which enable a proper disorder in the supercell structure, include potential ordering effects,\cite{Do2014} and crucially also exploring the effect of the exchange-correlation choice on the relative position of the impurity band. 

Previously, Aliev {\itshape et al.} reported a band gap value for ZrNiSn of \unit[190]{meV} by analysing the high temperature, intrinsic regime of their experimental resistivity data, assuming $\rho\propto\exp{-E_g/2k_BT}$.\cite{Aliev1989} 
This expression assumes Boltzmann statistics, which is requires that $E_F-E_C\gg k_BT$,
which we find not to be appropriate based on analysis using Fermi-Dirac statics, which results in $\Delta=\unit[40]{meV}$, i.e. $\lvert E_F-E_C\rvert\le kT$ in the studied temperature range.
In fact, in an Arrhenius plot, the resistivity of our study has a very similar slope as the data of Aliev {\itshape et al.}, see Fig. S5.

From the fitted $N_p$ values of our model, we can estimate the concentration of Ni interstitials in our samples. If each interstitial Ni atom contributed with four in-gap states (the $e_g$ orbitals), the fitted $N_\text{imp}\approx\unit[20\cdot10^{19}]{cm^{-3}}$ would correspond to a Ni interstitial concentration of \unit[$\sim1$]{\%}. 
This rough estimate compares well with values reported in the literature. 
For example, the measured curve of the Seebeck coefficient and resistivity for our ZrNiSn sample lies between the data of samples with nominal composition of ZrNiSn and ZrNi$_{1.01}$Sn reported by Romaka {\itshape{et al.}}.\cite{Romaka2013} Also the lattice parameter of our sample is close to the value reported for stoichiometric ZrNiSn \cite{Chauhan2019}, indicating again a low concentration ($\le 1 \%$) of Ni interstitials present in our samples.
However, further refinement of the estimated Ni interstitial concentration requires also additional insight into possible charge compensation and band width of the Ni interstitial impurity band. Band structure calculations indeed show a band width of ca. \unit[0.1]{eV} of the Ni interstitial impurity band, indicating a lifted degeneracy.\cite{Do2014,Douglas2014a}
We also note that the mobility ratio of electrons in the conduction band (\unit[$\sim30$]{cm$^2$V$^{-1}$s$^{-1}$, estimated from the high $T$ values in Fig.~\ref{fig_raw_data} (e), where $\sigma_\text{CB}\ll\sigma_\text{IB}$}
) to holes in the impurity valence band (\unit[$\sim4.5$]{cm$^2$V$^{-1}$s$^{-1}$}) obtained for our model is with $\sim6.7$ close to the ratio of 5 as suggested by Schmitt {\itshape et al.}. \cite{Schmitt2015}

Figure~\ref{fig_impurity_conc} shows some additional results obtained with our model:
(a) By choice, the impurity band is modelled as half filled at low temperatures. With increasing temperature, electrons are thermally excited into the conduction band and the population of the impurity band decreases, Fig. \ref{fig_impurity_conc} (a).
At low temperatures, the impurity band of ZrNiSn$_{0.9}$Pb$_{0.1}$ empties faster with increasing temperature because of $\Delta_\text{ZrNiSn$_{0.9}$Pb$_{0.1}$}<\Delta_\text{ZrNiSn}$. At higher temperatures, thermal excitation of electrons from the reservoir level into the impurity band of ZrNiSn$_{0.9}$Pb$_{0.1}$ reverses this trend and the impurity band of ZrNiSn empties faster. 
With help of the modeled results, we can further calculate the contribution of the impurity band to the electronic transport in ZrNiSn. The total conductivity is just the sum of the conductivity within the CB and within the impurity band, {\itshape cf.} Eq. \eqref{eq:cond_imp} and Figs. \ref{fig_impurity_conc} (c) and (d): At low temperatures, all electronic transport occurs within the impurity band, while the CB dominates the transport at higher temperatures.  $\sigma_\text{IB}$ still accounts for \unit[$\sim30$]{\%} of the total conductivity at room temperature.

\section{Summary}
In conclusion, we have shown that the transport properties of nominally undoped ZrNiSn and related compounds show signatures of impurity band conduction. By analyzing experimental results for the electrical resistivity, the Seebeck, Hall and Nernst coefficient with a semi-analytical model, we obtain excellent quantitative agreement with an impurity band located \unit[40]{meV} below the conduction band edge. A possible origin of the impurity band are interstitial Ni atoms, commonly found in these compounds.  
Our study should motivate further attempts to resolve the discrepancy between optical and transport band gaps, both experimental and theoretical. 
For example, one possibility is that excitations from the impurity band into the conduction band are optically forbidden or faint. 
Another explanation could involve different band features in different samples, caused by Ni-interstitial to clustering.\cite{Do2014} 
Optical measurements on samples with different Ni interstitial concentrations would thus be highly desirable, so would theoretical studies analyzing in detail the coupling of Ni interstitial states and the conduction band. 

\begin{acknowledgments}
This work was funded by the Research Council of Norway (THELMA 228854). MS gratefully acknowledges a traveling grant (No. 274164) from the Research Council of Norway for a three months stay at the Ohio State University.
Computations were performed on the Stallo high performance cluster through a NOTUR allocation.
\end{acknowledgments}

%

\end{document}